\documentclass[reqno,10pt,centertags,draft]{amsart}
\usepackage{amsmath,amsthm,amscd,amssymb,latexsym,upref}
\date{\today}

\newcommand{\bbD}{{\mathbb{D}}}
\newcommand{\bbI}{{\mathbb{I}}}

\newcommand{\bbR}{{\mathbb{R}}}

\newcommand{\bbZ}{{\mathbb{Z}}}
\newcommand{\bbC}{{\mathbb{C}}}

\newcommand{\bbT}{{\mathbb{T}}}

\newcommand{\cF}{{\mathcal{F}}}

\newcommand{\cG}{{\mathcal{G}}}

\newcommand{\cP}{{\mathcal{P}}}
\newcommand{\cQ}{{\mathcal{Q}}}

\newcommand{\cE}{{\mathcal{E}}}

\renewcommand{\Im}{\text{\rm Im}}
\renewcommand{\div}{\text{\rm div}}
\newcommand{\Jac}{\text{\rm Jac}}
\newcommand{\Julia}{\text{\rm Julia}}
\newcommand{\Drb}{\text{\rm Darb}}

\allowdisplaybreaks
\numberwithin{equation}{section}
\newtheorem{theorem}{Theorem}[section]

\newtheorem{lemma}[theorem]{Lemma}
\newtheorem{proposition}[theorem]{Proposition}
\newtheorem{corollary}[theorem]{Corollary}

\theoremstyle{definition}
\newtheorem{definition}[theorem]{Definition}

\newtheorem{conjecture}[theorem]{Conjecture}

\title[ Limit periodic Jacobi matrices
]{ Limit periodic Jacobi matrices
   with a singular continuous spectrum and the renormalization of
periodic matrices  }
\author{F. Peherstorfer, A. Volberg, P.Yuditskii}
   \thanks{Partially supported by NSF grant DMS-0200713,
   the Austrian Science
Fund FWF, project number: P16390--N04
and {\it Marie Curie International
Fellowship} within the 6-th European Community Framework Programme,
Contract Number: MIF1-CT-2005-006966}

\begin{document}

\begin{abstract}
For all  hyperbolic polynomials  we proved in \cite{VY}
a Lipschitz estimate of Jacobi matrices built by orthogonalizing
polynomials with respect to measures in the orbit of classical
Perron-Frobenius-Ruelle operators associated to hyperbolic polynomial
dynamics (with real Julia set). Here we prove that for  all
sufficiently hyperbolic polynomials this estimate becomes exponentially
better when the dimension of the Jacobi matrix grows. In fact, our main result
asserts
that a certain natural non-linear operator on Jacobi matrices built by a
hyperbolic polynomial with real Julia set is a contraction in operator norm if
the polynomial is sufficiently hyperbolic. This allows us to get for such
polynomials the solution of a problem of Bellissard, in other words, to prove
the limit periodicity of the limit Jacobi matrix. This fact does not
require the
iteration of the same fixed polynomial, and therefore it gives a wide class of
limit periodic Jacobi matrices with singular continuous spectrum.
\end{abstract}

\maketitle

\section{Introduction}
Let
$T$
be an expanding polynomial with the real Julia set $\Julia(T)$, $\deg T=
d$. We recall that $\Julia(T)$ is a nonempty compact set of
points which do not go to infinity under forward iterations of $T$.
Under the normalization
\begin{equation}
\label{norm}
T^{-1}:[-1,1]\to [-1,1];\,\,\, \pm 1\in \Julia(T)
\end{equation}
such a polynomial is well-defined by the position
of its critical values
$$
CV(T):=\{t_i=T(c_i): T'(c_i)=0,\  c_i>c_j \ \text{for}\ i>j
\}.
$$
Expanding, or hyperbolic polynomials are those, for which
$$
c_i\notin \Julia(T), \forall i\,,
$$
which is the same as to say that $CV(T)\cap \Julia(T)=\emptyset$ (just
use the fact that $\Julia(T)$ is invariant under taking full preimage
$T^{-1}$). The term ``expanding" is deserved because for expanding
polynomials one has the following inequality
\begin{equation}
\label{expanding}
\exists Q >1,\,\,|(T^n)'(x)| \geq c Q^n, \forall x\in \Julia(T)\,.
\end{equation}
Here and in everything what follows $T^n$ means $n$-th iteration of $T$,
$T^n = T\circ T\circ....T$.

Let us mention that for $T$ with a {\it real}
Julia set one has $|T(c_i)|>1$ since all solutions
of $T(x)=\pm 1$ should be real.

We will need to consider the notion of  ``sufficiently expanding"
(``sufficiently hyperbolic") polynomials. As we saw, the expanding
property is the
same (in our normalization) as $\text{dist}(CV(T), [-1,1])>0$. The polynomial
$T$ with normalization \eqref{norm} will be called {\it sufficiently
hyperbolic}
(or {\it sufficiently expanding}) if
\begin{equation}
\label{sh}
\text{dist}(CV(T), [-1,1]) \geq A\,,
\end{equation}
where $A$ is a large {\it absolute} constant to be specified later
(but $A=10$ will work).
Notice that the definition of sufficient hyperbolicity does not
involve the degree of $T$. In particular, $T$ and any of its iterative
powers $T^2, T^3,...$ are sufficiently hyperbolic simultaneously.

A Jacobi matrix $J:l^2(\bbZ)\to l^2(\bbZ)$ is called almost periodic if
the family
$$
\{S^{-k}JS^k\}_{k\in \bbZ},
$$
where $S$ is the shift operator in $l^2(\bbZ)$,
$S|m\rangle=|m+1\rangle$,
is a precompactum in the operator topology.

\smallskip

\noindent
{\bf Example}.
Let $G$ be a compact abelian group, $p(\alpha), q(\alpha)$ be continuous
functions on $G$,  $p(\alpha)\ge 0$. Then $J(\alpha)$ with the
coefficient sequences $\{p(\alpha+k\mu)\}_k, \{q(\alpha+k\mu)\}_k$,
$\mu\in G$, is almost periodic.

\smallskip

Let us show that in fact this is a general form of almost periodic
Jacobi matrices. For a given almost periodic $J$ define the metric on
$\bbZ$ by
$$
\rho_J(k):=||S^{-k}JS^k-J||.
$$
Evidently $\rho_J(k+m)\le\rho_J(k)+\rho_J(m)$.  Then $J=J(0)$,
where $G=I_J$, $I_J$ is the closure
of $\bbZ$ with respect to $\rho_J$, and $\mu=1\in I_J$.

Recall that for a given system of integers $\{d_k\}_{k=1}^\infty$
one can define
  the set
  \begin{equation}
\bbI=\underleftarrow{\lim}\{\bbZ/d_1...d_k\bbZ\},
\end{equation}
  that is $\alpha\in \bbI$ means that $\alpha$ is
  a sequence $\{\alpha_0,\alpha_1,\alpha_2,...\}$ such that
  $$
  \alpha_k\in \bbZ/d_1...d_{k+1}\bbZ\quad \text{and}
  \quad \alpha_k|\text{mod} \,d_1...d_k=\alpha_{k-1}.
  $$
The addition in $\bbI$ is defined as addition in
the $l$-th entry. The metric $\text{dist} (\alpha,
\beta) = \kappa^l$, where $\kappa\in (0,1)$, $l$ is the smallest integer
such that
$\alpha_l \neq\beta_l$, makes $\bbI$ a compact abelian group.
  In particular, if $p$ is a prime number and
  $d_k=p$ we get the ring of $p$--adic integers, $\bbI=\bbZ_p$.

In this work we build a certain machinery that allows
to construct almost periodic Jacobi matrices with singularly
continuous spectrum such that $I_J=\bbI$.

The key element of the construction is the following

\begin{theorem} \label{mainth}
Let $\tilde J$ be a Jacobi matrix
with the spectrum on $[-1,1]$. Then the following Renormalization
Equation has a
solution $J=J(\tilde J)=J(\tilde J; T)$ with the spectrum on $T^{-1}([-1,1])$:
\begin{equation}\label{3a}
V^*(z-J)^{-1}V=
(T(z)-\tilde J)^{-1}T'(z)/d,
\end{equation}
where $V|k\rangle= |dk\rangle$.
Moreover,
  if
$\min_i |t_i|\ge 10$ then
$$
||J(\tilde J_1)-J(\tilde J_2)||\le
\kappa ||\tilde J_1-\tilde J_2||.
$$
with an absolute constant $\kappa<1$ (does not depend on $T$ also).
\end{theorem}

This theorem, for example, will result in the following consequence:

\begin{theorem}
\label{lp}
Let $T$ be sufficiently hyperbolic in the sense of
\eqref{sh}.   Let $J_{\omega}$ be the Jacobi
matrix obtained by orthogonalizing polynomials with respect to the
balanced (equilibrium) measure
  $\omega$  on the Julia set of $T$.
Then $J_{\omega}$  is a limit periodic matrix. In other words, the
sequences that give the diagonal and the below (above) diagonal entries
are uniform limits of periodic sequences.
\end{theorem}

\bigskip

\noindent {\bf Remarks.} 1) A known problem
(due to Bellissard) asks to prove this statement for {\it all}
hyperbolic polynomials normalized as in \eqref{norm}.  Here we do it
for all sufficiently hyperbolic polynomials (recall that we always
tacitly assume that $\Julia(T)$ is on the real line). Our result about
sufficiently hyperbolic  polynomials explains, in a sense, the earlier
results in \cite{BGH}, \cite{BBM}, where it has been shown that
polynomials $T(z):=\varepsilon^{-n}\mathcal T_n(\varepsilon z)$ generate
limit periodic Jacobi matrices if $\varepsilon$ is sufficiently small,
here
$\mathcal T_n$ is the $n$-th Tchebyshef polynomial.  Smallness of
$\varepsilon$  obviously makes
$T=\varepsilon^{-n}\mathcal T_n(\varepsilon z)$
sufficiently hyperbolic in our sense \eqref{sh}.

\medskip

2) In the thesis of Herndon \cite{H}
Theorem \ref{lp} is proved by another method.
We regret that it has not
been published, that might have clarified the proof,
which seems to be
quite involved.

\medskip

3) One can wonder after analyzing the
results of \cite{BGH}, \cite{BBM} and the present paper, that may be
there is a threshold of hyperbolicity: before it $J_{\omega}$ is {\it
not} limit periodic, and after it it is limit periodic. However, we do
not believe in this sort of  behavior, but at this stage we cannot
prove the conjecture of Bellissard for {\it all} hyperbolic polynomial
with real Julia set.

\medskip

4) Let us mention that, in fact, \eqref{3a} has $2^{d-1}$ solutions
such that
the spectrum of $J$ is on $T^{-1}([-1,1])$. Here we use only one of them.

\bigskip

We note that the real output of Theorem \ref{mainth} is much wider than Theorem
\ref{lp}. It shows that

\begin{itemize}

\item
[a)] roughly speaking, constructing in a regular iterative way a Cantor
set $E$, $E\subset\dots\subset E_{n+1}\subset E_{n}\dots$, that may
support the spectrum of a limit--periodic Jacobi matrix it is enough to
follow the strategy:  on each step the approximating set
$E_n$ should have a form of an inverse polynomial image, i.e.:
$$
E_n=U_n^{-1}[-1,1], \quad U_n\ \text{ is a polynomial},
$$
or, what is
the same,
$E_n$ should be the spectrum of a periodic Jacobi matrix;

\item
[b)] the above statement becomes a theorem if on each step we remove
from the previous set a sufficiently large part (using sufficiently expanding
polynomials), i.e.: if $T_1, T_2...$, is a sequence of polynomials with
sufficiently large critical values, then $E_n:=U_n^{-1}[-1,1]$, with $U_n=
T_n\circ ...\circ T_2\circ T_1$;

\item [c)] the set $E$, that was constructed in the above
described way, is the spectral set of infinitely many (uncountable
set) of {\it different} limit periodic Jacobi matrices, that means
that each of the matrices does not belong to the hull of another
one (can't be obtained as a limit of shifts). The problem: to
describe the set of all limit periodic Jacobi matrices with
spectrum $E$ or certain subclasses (or at least to try to
understand how these sets look like), is a very interesting and
challenging problem.
\end{itemize}

Let us outline a proof of claim b). First,
we point out the following two properties
of the function $J(\tilde J; T)$ in Theorem \ref{mainth}. Due to the
commutant relation  $VS=S^d V$ one gets
${J(S^{-m}\tilde J S^m)=S^{-d m}J(\tilde J)S^{d m}}$.
The second  property is
$$
J(J(\tilde J; T_2); T_1)=
J(\tilde J; T_2\circ T_1),
$$
that is, the chain rule holds.

Now,
we produce  a limit periodic Jacobi matrix with singularly
continuous spectrum and such that $I_J=\bbI$.
For the  chosen system of polynomials
$T_1, T_2...$,  $\deg T_k=d_k$, with sufficiently large
critical values,
define
$J_m=J(\tilde J; T_m\circ ...\circ T_2\circ T_1)$.
By Theorem \ref{mainth}, the limit
$J=\lim_{m\to\infty} J_m$ exists and does not depend on $\tilde J$. Moreover,
$$
\forall j,\,\,||J-S^{-d_1...d_l j}JS^{d_1...d_l j}||\le
||J-S^{-j}JS^{j}||\,\kappa^l\le 2 \kappa^l.
$$
That is $\rho_J$  defines on $\bbZ$ the standard
$p$--adic topology  in this
case. This proves that $J$ is a limit periodic matrix,
in particular, it is almost periodic.

Notice that for the case $T_1=T_2=...=T_m=:T$,
  we  get  the limit periodic matrix with
the spectrum on  $\Julia(T)$.

\section{
Renormalization equation}

In this section
it is convenient to assume that
   $$
   T(z)=z^d-qdz^{d-1}+...
   $$
   is a {\it monic}
  expanding polynomial. Under this normalization
  $T^{-1}:[-\xi,\xi]\to [-\xi,\xi]$ for a certain $\xi>0$.

  Let $\tilde
J:l^2(\bbZ)\to l^2(\bbZ)$ be a Jacobi matrix with the spectrum on
$[-\xi,\xi]$. We
describe the set of solutions of the Renormalization Equation
\begin{equation}\label{t01}
V^*(z-J)^{-1}V=(T(z)-\tilde J)^{-1}T'(z)/d,
\quad V|k\rangle=|kd\rangle,
\end{equation}
here $J$ is a Jacobi matrix with the spectrum on $T^{-1}([-\xi,\xi])$
that should
satisfy \eqref{t01}.

In what follows  by $l^2_{\pm}(s)$ we denote the spaces
  which are formed by
$\{|s+k\rangle\}$ with $k\le 0$ and $k\ge 0$ respectively, that
is $l^2(\bbZ)=l^2_-(s)\oplus l^2_+(s+1)$. Correspondingly to these
decompositions we set
$\tilde J_{\pm}(s)=P_{l^2_{\pm}(s)}\tilde
J|l^2_{\pm}(s)$.

Recall that  a (finite or infinite)
one--sided Jacobi matrix is uniquely determined by its
so called resolvent function
\begin{equation}\label{4aug22}
\tilde r_{\pm}(z,s)=\langle s|(\tilde J_{\pm}(s)-z)^{-1}
|s\rangle,
\end{equation}
for which the following
decomposition in the continued fraction holds true
\begin{equation}\label{cf25av}
\tilde r_{+}(z,s)=
\cfrac{-1}{
z-\tilde q_s- \cfrac{\tilde p^2_{s+1}} {{z-\tilde
q_{s+1}-...}}}=
\frac{-1}{z-\tilde q_s+
\tilde p^2_{s+1}\tilde r_{+}(z,s+1)}.
\end{equation}

\begin{lemma}\label{1.1}
Assume that a matrix $J$ satisfies \eqref{t01}.
Then
\begin{equation}\label{4.1}
p_{sd+1}...p_{sd+d}=\tilde p_{s+1}
\quad\text{and}\quad q_{sd}=q.
\end{equation}
Further, let $J^{(s)}$ be the
$s$-th $d\times d$ block of  $J$, that is,
\begin{equation}\label{4aug23}
J^{(s)}=
\begin{bmatrix}
q_{sd}&p_{sd+1}& & & \\
p_{sd+1}&q_{sd+1}&p_{sd+2}& & \\
     &\ddots&\ddots&\ddots & \\
& & p_{sd+d-2}&q_{sd+d-2}&p_{sd+d-1} \\
&  & &p_{sd+d-1}&q_{sd+d-1}
\end{bmatrix}.
\end{equation}
Then its resolvent function is of the form
\begin{equation}\label{3}
\left<0\left|(z-J^{(s)})^{-1} \right|0\right>=
\frac{T'(z)/d}{T^{(s)}(z)},
\end{equation}
where $T^{(s)}(z)$ is a monic polynomial of degree $d$.
Moreover, at the critical points $\{c: T'(c)=0\}$ the following
recurrence relation holds
\begin{equation}\label{13aug23}
T^{(s)}(c)
+ \frac{\tilde p^2_s} {T^{(s-1)}(c)}=
T(c)-\tilde q_s.
\end{equation}
\end{lemma}

\begin{proof}
We write the Jacobi matrix $J$ as a $d\times d$ block matrix (each block
is of infinite size), that is, we are just reordering the standard basis:
\begin{equation}\label{5}
J=
\begin{bmatrix}
\cQ_{0}&\cP_{1}& & &S\cP_d  \\
\cP_{1}&\cQ_{1}&\cP_{2}& & \\
     &\ddots&\ddots&\ddots & \\
& & \cP_{d-2}&\cQ_{d-2}&\cP_{d-1} \\
\cP_d S^*&  & &\cP_{d-1}&\cQ_{d-1}
\end{bmatrix}.
\end{equation}
Here $\cP_k$ (respectively $\cQ_k$) is a diagonal matrix
$\cP_k=\text{diag}\{p_{k+sd}\}_{s\ge 0}$ and $S$ is the shift operator. With
respect to this reordering $V^*$ is the projection on the first--place
block--component.

Using this representation and the well
     known identity for  block matrices
\begin{equation*}
\begin{bmatrix}
A&B\\
C&D
\end{bmatrix}^{-1}=
\begin{bmatrix}
(A-B D^{-1}C)^{-1}&*\\
*&*
\end{bmatrix},
\end{equation*}
we get
\begin{equation}\label{6}
\frac{T(z)-\tilde J}{T'(z)/d}= z-\cQ_0-
\begin{bmatrix}\cP_1,&...,&S\cP_d
\end{bmatrix}
\{z-J_{1}\}^{-1}
\begin{bmatrix}
\cP_1\\
\vdots\\
\cP_d S^*
\end{bmatrix},
\end{equation}
where $J_{1}$ is the matrix that we obtain from $J$ by deleting
the first block--row and the first block--column in \eqref{5}.
Thus the second relation in \eqref{4.1} is already proved, we just
compare the leading terms in the decomposition over powers
of $1/z$ in the right and left hand sides and note that
the third term on the right is of  order $1/z$.

But the most important remark is that in $(z-J_{1})$ each block is a
diagonal matrix (means all diagonals are the {\it main} diagonals in each
block, on the contrary to
$J$ that contains $S\cP_d$ and $\cP_d S^*$).  That's why we can easily
get an inverse matrix in terms of the {\it scalar} orthogonal
polynomials.

Let us introduce the following notations: everything related to
$J^{(s)}$ has superscript $s$. For instance: $p_k^{(s)}=p_{sd+k}$,
$1\le k\le d$, respectively $P^{(s)}_d$ and $Q^{(s)}_d$ mean
orthonormal polynomials of the first and second kind, in particular,
\begin{equation}\label{rfaug24}
\langle 0|(z-J^{(s)})^{-1}|0\rangle=
\frac{Q^{(s)}_d(z)}{P_d^{(s)}(z)}.
\end{equation}

Let $J^{(s)}_1$ denote the matrix that we obtain from
$J^{(s)}$ (see \eqref{4aug23}) by deleting the first row and the first
column. Then, for  $J^{(s)}_1$, $Q^{(s)}_d$ are orthogonal polynomials
of the first kind and we denote by $R^{(s)}_d$
corresponding to them orthogonal polynomials of the
second kind. Note that $P^{(s)}_d$ and $R^{(s)}_d$ are related by
\begin{equation}\label{7aug23}
\frac{P_d^{(s)}(z)}
{Q^{(s)}_d(z)}=z-q_0^{(s)}- (p^{(s)}_{1})^2\frac{R_{d}^{(s)}(z)}
{Q^{(s)}_d(z)}.
\end{equation}
In these terms the four interesting for us elements
of the resolvent of $J^{(s)}_1$ are:
\begin{equation}\label{9aug23}
(z-J^{(s)}_1)^{-1}=
\begin{bmatrix}
\frac{R^{(s)}_d}{Q^{(s)}_d}&\dots
&\frac{1}{p^{(s)}_1 p^{(s)}_d Q^{(s)}_d}
\\
\vdots & & \vdots\\
\frac{1}{p^{(s)}_1 p^{(s)}_d Q^{(s)}_d} &\dots &\frac{
Q^{(s)}_{d-1}/p^{(s)}_d} {Q^{(s)}_d}
\end{bmatrix}.
\end{equation}

Recalling again that $J_1$ is just a block-diagonal matrix we
substitute \eqref{9aug23} in \eqref{6}.
As result in the RHS (as well as in the LHS) we get a three-diagonal
matrix. On the main diagonal we have
$$
z-\cQ_0-\cP_1\{(z-J_{1})^{-1}\}_{1,1}\cP_1-
S\cP_d\{(z-J_{1})^{-1}\}_{d-1,d-1}
\cP_d S^*,
$$
and each entry on the diagonal, due to \eqref{9aug23}
and then \eqref{7aug23}, is
\begin{equation*}
\begin{split}
z-q_0^{(s+1)}- (p^{(s+1)}_{1})^2\frac{R_{d}^{(s+1)}(z)}
{Q^{(s+1)}_d(z)}-&(p^{(s)}_{d})^2\frac{Q_{d-1}^{(s)}(z)/p^{(s)}_{d}}
{Q^{(s)}_d(z)}\\
= \frac{P_d^{(s+1)}(z)}
{Q^{(s+1)}_d(z)}- &(p^{(s)}_{d})^2\frac{Q_{d-1}^{(s)}(z)/p^{(s)}_{ds}}
{Q^{(s)}_d(z)}.
\end{split}
\end{equation*}
Comparing this with the LHS \eqref{6} we get
\begin{equation}\label{7}
\frac{T(z)-\tilde q_{s+1}}{T'(z)/d}= \frac{P_d^{(s+1)}(z)}
{Q^{(s+1)}_d(z)}- (p^{(s)}_{d})^2\frac{Q_{d-1}^{(s)}(z)/p^{(s)}_{d}}
{Q^{(s)}_d(z)}.
\end{equation}
Similarly, below the main diagonal on the right we have
$$
-S\cP_d\{(z-J_{1})^{-1}\}_{d-1,1}
\cP_1.
$$
  So, using \eqref{9aug23}, we get from \eqref{6}
\begin{equation}\label{8}
\frac{\tilde p_{s+1}}{T'(z)/d}= \frac 1{Q^{(s)}_d(z)}=
\frac{p_1^{(s)} ...p_d^{(s)}} {z^{d-1}+...}.
\end{equation}
Thus the first relation in \eqref{4.1} is also proved, moreover
all $Q^{(s)}_d(z)$ (independently on $s$), being normalized
to the leading coefficient one, coincides with $T'(z)/d$.

We define $T^{(s)}(z)=z^d+...$ by the same normalization
  \begin{equation}\label{2.15add}
  T^{(s)}(z):=\tilde p_{s+1} P^{(s)}_d(z).
  \end{equation}
Then \eqref{rfaug24} implies \eqref{3}.

Now we use the (last) well known fact on
orthogonal polynomials ---
the Wronskian identity:
$$
p_d^{(s)}Q^{(s)}_{d}(z)P^{(s)}_{d-1}(z)-
p_d^{(s)}P^{(s)}_{d}(z)Q^{(s)}_{d-1}(z)=1,
$$
due to which, if $T'(c)=0$ then
\begin{equation}\label{9}
-p^{(s)}_{d}{Q^{(s)}_{d-1}(c)} =\frac 1{P^{(s)}_d(c)}.
\end{equation}
So, combining  \eqref{9} with \eqref{8},  we get
from \eqref{7}
the recurrence relation
\begin{equation}\label{2.16}
T(c)-\tilde q_{s+1}= T^{(s+1)}(c)+\frac{\tilde p^2_{s+1}}
{T^{(s)}(c)}.
\end{equation}
Thus the lemma is completely proved.
\end{proof}

Now we are in a position to show that the Renormalization Equation
has $2^{d-1}$ solutions.
Then we show that
they are the only possible solutions.
This set of solutions we parametrize
by a collections of vectors
\begin{equation}\label{delta}
\delta:=\{\delta_c\}_c,
\end{equation}
where each component $\delta_c$ can be chosen as plus or
minus one.

\begin{theorem}\label{exs} Fix a vector $\delta$ of the form
\eqref{delta}.
For a given $\tilde J$ with the spectrum on $[-\xi,\xi]$ define the
Jacobi matrix $J$
according to the following algorithm:

For $s\in \bbZ$ we put
\begin{equation}\label{4m}
\frac 1{T^{(s)}(c)}=-\tilde r_-(T(c),s),
\quad\text{if}\  \delta_c=-1,
\end{equation}
  and
\begin{equation}\label{4p}
{T^{(s)}(c)}=-\tilde p_{s+1}^2\tilde r_+(T(c),s+1),
\quad\text{if}\  \delta_c=1,
\end{equation}
  where the functions $\tilde r_\pm(z,s)$
are defined by \eqref{4aug22}. Then define the monic polynomial
$T^{(s)}(z)$ of degree $d$ by the interpolation formula
\begin{equation}\label{4.2}
T^{(s)}(z)=(z-q)T'(z)/d+
\sum_{c:T'(c)=0}\frac{T'(z)}{(z-c)T''(c)}T^{(s)}(c).
\end{equation}
Define the block $J^{(s)}$ (see \eqref{4aug23}) by its resolvent
function according to \eqref{3}. Finally define the entry $p_{sd+d}$ by
\eqref{4.1}.

We claim that the matrix $J=J(\delta, \tilde J)$, combined with such
blocks and entries over all $s$, satisfies \eqref{t01}.
\end{theorem}

\noindent
{\bf Example.} The solution related to the vector
$$
\delta_-=\{-1,\dots ,-1\},
$$
that is all $T^{(s)}(c)$ are defined by \eqref{4m},
  plays the most important role in what follows. Precisely for this solution
  $J(\tilde J):=J(\delta_-,\tilde J)$ we prove the contractibility property (our
  main Theorem \ref{mainth}).

\begin{proof}
First of all let us mention that for all $c$, $|T(c)|>\xi$, that
is  ${T^{(s)}(c)}$ is well defined by either \eqref{4m}
or \eqref{4p}, moreover this value is of the same sign as
$T(c)$. It means that the rational function
$-\frac{T'(z)/d}{T^{(s)}(z)}$ is the Stieltjes transform of
a positive measure (supported at the zeros of $T^{(s)}(z)$), and hence
there exists a unique  $d\times d$ Jacobi matrix defined
by \eqref{4aug23}. Note that \eqref{4.2} implies
immediately that $q_{sd}=q$, again we look at the leading
term in the decomposition of the resolvent function into
the continued fraction.

Now, we see that \eqref{8} holds, it's just a matter
of the definition of $p^{(s)}_d$. We have to check \eqref{7}.
Using the form of $Q^{(s)}_d(z)$ we have equivalently
\begin{equation}\label{21aug25}
T(z)-\tilde q_{s+1}=T^{(s)}(z) +  \tilde
p_{s+1}p^{(s)}_{d}Q^{(s)}_{d-1}(z).
\end{equation}
Subtracting $(z-q)T'(z)/d$ from both parts we arrive at the question of the
identity of two polynomials of degree $d-2$. Thus, we need to check
\eqref{21aug25} only at the critical points. Using the Wronskian identity
\eqref{9} we get \eqref{2.16}. Of course, the main point is that
${T^{(s)}(c)}$,
being defined by either \eqref{4m} or \eqref{4p}, satisfy the recursion
\eqref{13aug23} (this is the continued fraction decomposition for $\tilde
r_{\pm}(z,s)$, see \eqref{cf25av}).

Thus having \eqref{6}, we proved \eqref{t01}.

\end{proof}

\begin{theorem}\label{uni}
Theorem \ref{exs} describes the whole set of solutions of the
Renormalization Equation.
\end{theorem}

\begin{proof}
We need to show that \eqref{4m}, \eqref{4p} give the complete possible
choice of the values $T^{(s)}(c)$, say for $s=0$. Then all other
values are of the same form due to Lemma \ref{1.1}.

We claim that any other choice of $T^{(0)}(c)$ contradicts to the
regularity of the resolvent of $J$ at $c$.

Using standard formulas for orthogonal polynomials for two--sided
Jacobi matrices (see Appendix, Corollary \ref{cor4.2}) we have from the
Renormalization  Equation
\begin{equation}\label{re0}
\begin{split}
\tilde r_-^{-1}(T(z),0)=&
r_-^{-1}(z,0) T'(z)/d+p^2_1\tilde p_1 R^{(0)}_d(z),
\\
\tilde p_1^2\tilde r_+(T(z),1)=&
p_1^2 r_+(z,1) T'(z)/d+p^2_1\tilde p_1 R^{(0)}_d(z),
\end{split}
\end{equation}
where $r_{\pm}(z,s)$ are the resolvent functions of
$J_{\pm}(s)$.

In the same time both functions
  $r_+(z,1)$ and $r_-^{-1}(z,0)$
can not have a pole at $c$ simultaneously.
This contradicts to (see \eqref{ap4.5})
\begin{equation}\label{res0}
\langle 1 |(J-z)^{-1}| 1\rangle={r_+(z,1)}
{r^{-1}_-(z,0)}
\langle 0 |(J-z)^{-1}| 0\rangle:
\end{equation}
$\langle 0 |(J-z)^{-1}| 0\rangle$ can not have zero multiplicity more
than one and $\langle 1 |(J-z)^{-1}| 1\rangle$ should be regular
at $c$.

By \eqref{7aug23} $T^{(0)}(c)=-p^2_1\tilde p_1 R^{(0)}_d(c)$,
so we get \eqref{4m} from \eqref{re0} if
${r^{-1}_-(z,0)}$ is regular and \eqref{4p} in the case when
$r_+(z,1)$ is regular at $c$.
\end{proof}

\noindent
{\bf Remark.} Formulas \eqref{4m}, \eqref{4p}
play the main role in solving the Renormalization Equation.
Actually, we proved or found them in Theorem \ref{uni} for $s=0$,
or, in the same way, for  any other fixed $s$. However, there is
another important ingredient: it should be also shown that if we choose
this or that definition of $T^{(s)}(c)$ for any particular $s$ we have to
  be consistent with this definition for all other values of $s$, that is
we can not define, for instance,
$T^{(0)}(c)$ by \eqref{4m} and $T^{(1)}(c)$ by
\eqref{4p} for the same $c$. Precisely this claim is the main output of
Lemma \ref{1.1}. \qed

Let us mention that the Renormalization Equation can be rewritten
equivalently in the  form of polynomials equations.

\begin{lemma} Equation \eqref{t01} is equivalent to
\begin{equation}
\label{re.1}
V^*T(J)=\tilde JV^*,
\end{equation}
\begin{equation}
\label{re.2}
V^*\frac{T(z)-T(J)}{z-J}V=T'(z)/d.
\end{equation}
\end{lemma}

\begin{proof}
Starting with \eqref{re.1}, \eqref{re.2} we get
\begin{equation*}
(T(z)-\tilde J)V^*(z-J)^{-1}V=
V^*\{T(z)-T(J)\}\{z-J\}^{-1}V=T'(z)/d.
\end{equation*}

Having \eqref{t01} we get
\begin{equation}\label{re.p.1}
\begin{split}
V^*\frac{T(z)-T(J)}{z-J}V=&
T(z)V^*(z-J)^{-1}V-V^*\frac{T(J)}{z-J}V\\
=&T(z)\frac{T'(z)/d}{T(z)-\tilde J}-V^*\frac{T(J)}{z-J}V\\
=&{T'(z)/d}+\tilde JV^*(z-J)^{-1}V-V^*\frac{T(J)}{z-J}V.
\end{split}
\end{equation}
Since the left hand side in \eqref{re.p.1} is a polynomial of $z$
we obtain two relations
\begin{equation*}
V^*\frac{T(z)-T(J)}{z-J}V=T'(z)/d
\end{equation*}
and
\begin{equation*}
\{\tilde JV^*-V^*T(J)\}\{(z-J)^{-1}V\}=0.
\end{equation*}
Since vectors of the form $(z-J)^{-1}V f$, $f\in l^2$, are complete
in $l^2$ the last relation implies \eqref{re.2}.
\end{proof}

\section{Proof of the main theorem}

We start with
\begin{lemma} \label{lemma3.1}
Let $J^{(s)}$ be the $s$--the block
of $J=J(\delta,\tilde J)$. Define the measure $\sigma^{(s)}$ by
\begin{equation}\label{3.1aug}
\frac{T^{(s)}(z)}{T'(z)/d}=z-q-\int\frac{d\sigma^{(s)}(x)}{x-z},
\end{equation}
that is, $\sigma^{(s)}$ is the spectral measure of
the obliterated matrix $J^{(s)}_1$ normalized by
$$
\int\, d\sigma^{(s)}(x)= (p^{(s)}_1)^2.
$$
Then
\begin{equation}\label{3.2aug}
\int\frac{d\sigma^{(s)}}{T^{(s)}(c)^2}=
\frac { p^2_{(s+1)d}}{\tilde p^2_{s+1}}.
\end{equation}
\end{lemma}
\begin{proof}
Note that $Q^{(s)}_{d-1}$ is the orthonormal
polynomial with respect to $\sigma^{(s)}$,
$$
\int (Q^{(s)}_{d-1}(c))^2 d\sigma^{(s)}(c)= 1.
$$
Using \eqref{9} and the normalization \eqref{2.15add}
we get \eqref{3.2aug}.
\end{proof}

The proof of the theorem is based on the following well--known and
simple lemma.

\begin{lemma}\label{3.1}
Assume that two non--normalized measures $\sigma$ and $\hat\sigma$
are mutually absolutely continuous. Moreover, that $d\hat\sigma
=f\,d\sigma$ and
$(1+\epsilon)^{-1}\le f\le (1+\epsilon)$. Let us associate with these measures
Jacobi matrices $J= J(\sigma)$, $\hat J= J(\hat\sigma)$. Then for their
coefficients we have
$$
|\hat p_s-p_s|\le \epsilon|| J||, \quad s\ge 0.
$$
\end{lemma}

\begin{proof} Assume that $p_s\ge\hat p_s$.
Let us use an extremal property of orthogonal polynomials,
\begin{equation*}
\begin{split}
(1+\epsilon)\hat p_0^2...\hat p_s^2=& (1+\epsilon)\int\hat
p_0^2...\hat p_s^2
\hat P_s^2\,d\hat\sigma \ge
\int\{z^s+...\}^2\,d \sigma\\
\ge& \inf_{\{P=z^s+...\}}\int P^2 \,d \sigma
= p_0^2... p_s^2.
\end{split}
\end{equation*}
Similarly
\begin{equation*}
(1+\epsilon)p_0^2... p_{s-1}^2
     \ge
\hat p_0^2...\hat p_{s-1}^2.
\end{equation*}
Therefore
$$
  p_s^2\le \hat p_s^2\le{(1+\epsilon)^2}p_s^2
$$
     and hence
$$
0\le \hat p_s-p_s\le \epsilon p_s.
$$

\end{proof}

\begin{proof}[Proof of  Theorem \ref{mainth}]
Given $\tilde J_{I}$ and $\tilde J_{II}$ let us compare the blocks
$J^{(s)}_{I}$ and $J^{(s)}_{II}$ of the matrices
$J_{I}:=J(\delta_-, \tilde J_{I})$ and
$J_{II}:=J(\delta_-, \tilde J_{II})$.
Actually we will apply Lemma \ref{3.1}
to  non--normalized spectral measures
$\sigma^{(s)}_{I}$ and
$\sigma^{(s)}_{II}$, see \eqref{3.1aug},
corresponding to the obliterated matrices $(J^{(s)}_{I})_1$
and $(J^{(s)}_{II})_1$.

Note that both measures are supported on the critical
points $\{c: T'(c)=0\}$ and, therefore, they are mutually absolutely
continuous. Moreover, the density of the second measure
with respect to the first one is of the form
$$
f(c)=f^{(s)}(c):=\frac{T^{(s)}_{II}(c)}{T^{(s)}_{I}(c)}.
$$
Assuming $f(c)\ge 1$ let us estimate $f(c)-1$ from above.

\begin{equation}
\begin{split}
f(c)-1=&\frac{1/T^{(s)}_{I}(c)-1/T^{(s)}_{II}(c)}{1/T^{(s)}_{II}(c)}\\
=&
\frac{\langle s|(T(c)-\tilde J_{II,-}(s))^{-1}
(\tilde J_{I,-}(s)-\tilde J_{II,-}(s))(T(c)-\tilde
J_{I,-}(s))^{-1}|s\rangle}{1/T^{(s)}_{II}(c)}.
\end{split}
\end{equation}
Since the spectrum of $\tilde J_{II,-}(s)$ is on $[-\xi,\xi]$
we get, by definition \eqref{4m},
$$
\left|\frac{1}{{T^{(s)}_{II}(c)}}\right|
\ge\frac 1{|T(c)|+\xi},
$$
and, for the same reason,
$$
||
(T(c)-\tilde J_{i,-})^{-1}|s\rangle
||\le\frac 1{|T(c)|-\xi}, \ i=I \ \text{or} \ i= II.
$$
Therefore
\begin{equation}\label{ef}
0\le f(c)-1\le ||\tilde J_{I}-\tilde J_{II}||\frac
{|T(c)|+\xi}{(|T(c)|-\xi)^2}.
\end{equation}
Thus, by Lemma \ref{3.1}, we obtain
\begin{equation}\label{e1}
|(p_{I})_{sd+k}-(p_{II})_{sd+k}|\le\varkappa ||\tilde J_{I}-\tilde
J_{II}||,
\quad 1\le k\le d-1,
\end{equation}
where
$$
\varkappa:=\max_{c}\frac
{|T(c)|/\xi+1}{(|T(c)|/\xi-1)^2}.
$$

We have to estimate $|(p_{I})_{sd+d}-(p_{II})_{sd+d}|$. Note that due to
  \eqref{4.1} and Lemma \ref{lemma3.1}
\begin{equation}\label{14}
\frac{1}{ (p_{i})_{sd+1}...(p_{i})_{sd+d-1}}=
\frac { (p_i)_{(s+1)d}}{\tilde p_{s+1}}
\le\frac 1{|T(c)|/\xi-1}, \ i=I \ \text{or} \ i= II.
\end{equation}
Now,
\begin{equation*}
\begin{split}
(p_I)_{sd+d}-(p_{II})_{sd+d}=&
\frac{(\tilde p_I)_{s+1}}{ (p_I)_{sd+1}...(p_I)_{sd+d-1}}-
\frac{(\tilde p_{II})_{s+1}}{ (p_{II})_{sd+1}...(p_{II})_{sd+d-1}}\\
=&
\frac{(\tilde p_I)_{s+1}-
(\tilde p_{II})_{s+1}}{ (p_I)_{sd+1}...(p_I)_{sd+d-1}}\\
+&\frac{(\tilde p_{II})_{s+1}}{ (p_I)_{sd+1}...(p_I)_{sd+d-1}}
\left(1-
\frac{ (p_I)_{sd+1}...(p_I)_{sd+d-1}}
{ (p_{II})_{sd+1}...(p_{II})_{sd+d-1}}
\right)
\end{split}
\end{equation*}
Using \eqref{14}, \eqref{ef} and Lemma \ref{3.1}
we obtain
\begin{equation}\label{e2}
\begin{split}
|(p_I)_{sd+d}-(p_{II})_{sd+d}|\le&
||\tilde J_1-\tilde J_2||\max_c\frac 1{|T(c)|/\xi-1}\\
+&\max_c\frac 1{|T(c)|/\xi-1}||\tilde J_1-\tilde J_2||\varkappa/2\\
=&||\tilde J_1-\tilde J_2||\max_c\frac 1{|T(c)|/\xi-1}
(1+\varkappa/2).
\end{split}
\end{equation}
Thus \eqref{e1} and \eqref{e2} show that say for
$\min_c{|T(c)|/\xi}\ge 10$ the renormalization is a contraction.
\end{proof}

\section{Proof of Theorem \ref{lp} }

\begin{proof}
For a given sufficiently hyperbolic polynomial $T$ we define
$$
J_{n+1}=J(\delta_-, J_n)
$$
starting from an arbitrary initial $J_0=\tilde J$ with the spectrum
on $[-\xi,\xi]$.
Due to the contractibility of the renormalization $J_n$ converges in the
operator norm to $J$, moreover $J$ has the spectrum on $\Julia(T)$
and it is limit
periodic
$$
\Vert J-S^{-d^nl}JS^{d^nl}\Vert\le 2\xi \kappa^n,\ \kappa<1.
$$

We claim that $J$ is an orthogonal sum of two one-sided Jacobi matrices
\begin{equation}\label{4.1}
J=\begin{bmatrix} J_-(-1)& 0\\
0&J_+(0)
\end{bmatrix}.
\end{equation}
That is, we claim that $p_0=0$. Indeed, by Lemma \ref{lemma3.1} we have
\begin{equation}\label{4.2}
p(n+1)_{sd}\le \left\{\max_{c}\frac{1}{|T(c)|/\xi-1}\right\} p(n)_s\le
\kappa p(n)_s,
\end{equation}
where $p(n)_s$ is the $s$--th coefficient in the matrix
$J_n$. Therefore, all $s d^l$--th coefficients of $J$ are uniformly small
$$
p_{s d^l}\le 2\xi \kappa^l,
$$
and, in particular, $p_0=0$.

Thus $J_+:= J_+(0)$ is a one--sided Jacobi
limit--periodic
matrix with the spectrum on $\Julia(T)$, moreover, its spectral
measure $\sigma_+$
(supported on $\Julia(T)$) possesses the renormalization property
\begin{equation}\label{4.3}
\int \frac{T'(z)/d}{T(z)-x}d\sigma_+(x)=
\int \frac{1}{z-x}d\sigma_+(x).
\end{equation}
This means that $\sigma_+$ is an eigen--measure for the
Ruelle operator $L^*\sigma_+=\sigma_+$, where the operator $L$
acts on a continuous  function $f$ on $\Julia(T)$ by
\begin{equation}\label{is4.4}
    (L f)(x)=\frac 1 d\sum_{T(y)=x} f(y).
\end{equation}
In other words $\sigma_+$ is the balanced measure on $\Julia(T)$.

\end{proof}

Note that due to the Renormalization Equation the spectral measure of
$J_-$ is the eigen--measure for the Ruelle operator
\begin{equation}\label{is4.5}
(L_2 f)(x)=\sum_{T(y)=x} \frac{f(y)}{T'(y)^2},
\end{equation}
i.e. $L^*_2\sigma_-=\rho\sigma_-$, $\rho>0$. In the case of quadratic
polynomials this fact was proved in \cite{SYu}.

\section{
The renormalization
of periodic matrices}

The renormalization \eqref{t01} acts in the most
natural way on periodic Jacobi matrices. We recall some basic facts from the
spectral
theory of such matrices.

The spectrum $E$ of any periodic matrix $J$ is an inverse polynomial
image
\begin{equation}\label{4.4}
E=U^{-1}[-1,1]
\end{equation}
the polynomial $U$ of degree $g+1$ should have
all critical points $\{c_U\}$ real and for all critical values
$|U(c_U)|\ge 1$. For simplicity we will assume $|U(c_U)|> 1$.
Then the spectrum of $J$ consists of $g$ intervals
$$
E=[b_0,a_0]\setminus (\cup_{j=1}^g(a_j,b_j)).
$$
Also it would be convenient for us to normalize $U$ by a linear
change of the variable such that $b_0=-1$ and $a_0=1$.

Having the set $E$ of the above form fixed, let us describe the whole set
of periodic Jacobi matrices $J(E)$ with the given spectrum.
To this end we associate with $U$ the hyper--elliptic Riemann surface
$$
X=\{Z=(z,\lambda): \lambda^2-2 U(z)\lambda+1=0\}.
$$
The involution on it we denote by $\tau$,
\begin{equation}\label{invol}
\tau Z:=\left(z,\frac{1}{\lambda}\right)\in X.
\end{equation}
The set
$$
X_+=\{Z\in X: |\lambda(Z)|<1\}
$$
we call the upper sheet of $X$. Note $X_+\simeq \bar \bbC\setminus E$,
in fact, $z(Z)\in \bar \bbC\setminus E$ if $Z\in X_+$.

The following well known theorem describes $J(E)$ in terms of {\it real}
divisors on
$X$. The Jacobian variety of $X$, $\Jac(X)$, is a $g$ dimensional
complex torus,
$\Jac(X)\simeq \bbC^g/L(X)$, where $L$ is a lattice (that can be chosen in the
form $L=\bbZ^g+\Omega \bbZ^g$ with $\Im \Omega>0$). Consider the $g$
dimensional
real subtorus consisting of divisors of the form
$$
D(E)=\{D=D_+ -D_C,\ D_+:= \sum_{i=1}^g Z_i: Z_i\in X,
\ z(Z_i)\in[a_i,b_i]\},
$$
here $D_C$ is a point of normalization that we choose of the form
$$
D_C:=\sum_{i=1}^g C_i
: C_i\in X,
\ z(C_i)=(c_U)_i, |\lambda(C_i)|>1,
$$
--- the collections of the points on the lower sheet with the
$z$--coordinates at the critical points. (At least topologically,
it is evident $D(E)\simeq \bbR^g/\bbZ^g$).

\begin{theorem}
For given $E$ of the form \eqref{4.4} there exists an one--to--one
correspondence
between $J(E)$ and $D(E)$.
\end{theorem}

Let now $\tilde U$ be a polynomial of the above described form, we restore the
normalization $T^{-1}:[-1,1]\to [-1,1]$ for the expanding polynomial, and we
define $U=\tilde U\circ T$. Then we have a covering $\pi$ of the
Riemann surface
$\tilde X$ associated to $\tilde U$ by the surface $X$ associated to $U$:
\begin{equation}\label{cov}
\pi(z,\lambda)=(T(z), \lambda),
\end{equation}
note $\pi: X_+\to \tilde X_+$.

According to the general theory, this covering generates different natural
mappings \cite{Mam},
in particular,
\begin{equation}\label{covjac}
\pi_*: \Jac(X)\to \Jac(\tilde X),
\end{equation}
and
\begin{equation}\label{covjacback}
\pi^*: \Jac(\tilde X)\to \Jac(X).
\end{equation}
In this section we solve equations \eqref{re.1}, \eqref{re.2}
using this language, see Theorem \ref{th5.8}.  Note that  \eqref{re.1}
already guarantied that $J\in J(E)$ implies $\tilde J\in J(\tilde E)$.

\bigskip

To continue  we need to recall some special functions
on hyper--elliptic Riemann surfaces.

The first object is the Complex Green's function.
Note that the function
$\lambda$ in
$X_+$ has no zeros except for infinity, where it has a zero of
multiplicity
$g+1$, moreover $|\lambda|=1$ on $\partial X_+$. We define the Complex
Green's function (with respect to infinity) by $b^{g+1}=\lambda$. It is not
single valued in $X_+$ but it has the only {\it simple} zero at infinity. Note
that
$$
G(z)=\log\frac{1}{|b(z)|},
$$
where $G(z)=G(z,\infty)$ is the standard Green's function for the domain $\bar
\bbC\setminus E$. Generally,
$$
G(z,z_0)=\log\frac{1}{|b_{z_0}(z)|},
$$
defines the Complex Green's function $b_{z_0}$ with the only zero at $Z_0\in
X_+$,
$z(Z_0)=z_0\in \bar \bbC\setminus E$.

Since $\tilde \lambda\circ \pi= \lambda$ we have the relation
\begin{equation}\label{4.7s}
\tilde b\circ\pi= b^d
\end{equation}

The differential $ \frac{1}{2\pi i}d\log b$,
being restricted
on $\partial X_+$, is the  harmonic measure $d\omega$ of the domain
$\bar
\bbC\setminus E$
with pole at infinity.

  The  space $L^p(\partial X_+)$, in a sense, is the $L^p$
space with respect to the harmonic measure, but it should be mentioned
that $\partial X_+= (E-i0)\cup(E+i0)$, i.e.,
an element $f$ of $L^p(\partial X_+)$ may have different values
$f(x+i0)$ and $f(x-i0)$, $x\in E$.

Having in mind \eqref{4.7s} we get
\begin{equation}\label{4.8s}
\int_{\partial X_+} f\, d\omega
= \int_{\partial \tilde X_+} \frac 1 d
\left(\sum_{\pi( Z)= \tilde Z}f(Z) \right)(\tilde Z)\, d\tilde\omega
\end{equation}
for every $f \in L^1(\partial X_+)$.

\begin{definition}\label{def4.2}
The Hardy space $H^2(X_+)$ consists of functions $f$
holomorphic on $X_+$ (or what is the same in the domain
$\bar \bbC\setminus E$) having harmonic majorant
\begin{equation}\label{4.5}
|f(z)|^2\le u(z),\ z\in \bar \bbC\setminus E,
\end{equation}
where $u(z)$ is harmonic in $\bar \bbC\setminus E$. The norm of $f$ is
defined by
$$
\Vert f\Vert^2:=\inf_{u} u(\infty),
$$
where $u$ runs over all harmonic  functions
satisfying \eqref{4.5}
\end{definition}

An equivalent way to define $H^2(X_+)$ is to close the set of
holomorphic functions uniformly bounded in $ X_+$ with respect to the norm
\begin{equation}\label{4.6}
\Vert f\Vert^2:=\int_{\partial X_+}|f|^2\, d\omega.
\end{equation}

As it follows directly from \eqref{4.8s},
the covering \eqref{cov} generates an isometrical enclosure
\begin{equation}\label{encl1}
v_+: H^2(\tilde X) \to H^2(X_+)
\end{equation}
acting in a natural way
\begin{equation}\label{encl2}
(v_+f)(Z)= f(\pi(Z)).
\end{equation}

Now we have to describe the most complicated but the most
important element of the construction: we have to introduce a very
natural orthonormal basis in $H^2(X_+)$.
The multiplication operator by $z$, with respect to this basis, will lead
us to Jacobi matrices, the substitution \eqref{encl2} to the
isometry $V$ and so on...
  This basis  is a counterpart of the standard basis of
$\{\zeta^n\}_{n\ge 0}$ in the standard Hardy space $H^2(\bbD)$, $\bbD
=\{|\zeta|<1\}$.

Note that $1\in  H^2(X_+)$, moreover,
$$
\langle f,1\rangle = f(\infty)
$$
for every $f\in  H^2(X_+)$. Therefore the orthogonal complement to $1$
consists of functions with $f(\infty)=0$. Let us give an alternative
description of
$$
H^2_0(X_+)=\{f\in H^2(X_+): f(\infty)=0\}.
$$

Any function from $H^2_0(X_+)$, having zero at infinity, is the form
$f= b\hat f$. However $b$ is not single--valued, thus so is $\hat f$.
We need to
generalize slightly Definition \ref{def4.2}.

\begin{definition}
Let $\Gamma=\Gamma(E)$ be the fundamental group of the domain $\bar
\bbC\setminus E$. Let $\alpha$ be an element of the dual group of characters
$\Gamma^*$, that is, for any contour $\gamma\in \Gamma$ in the domain,
$\gamma\mapsto\alpha(\gamma)$, where $\alpha(\gamma)$ is a number of
absolute value one, and for any two contours $\gamma_1$, $\gamma_2$
$$
\alpha(\gamma_1\gamma_2)=\alpha(\gamma_1)\alpha(\gamma_2).
$$
The Hardy space $H^\infty(X_+,\alpha)$ consists of
holomorphic multivalued functions $f$ uniformly bounded in the domain
$\bar \bbC\setminus E$ such that
$$
f(\gamma z)=\alpha(\gamma) f(z),
$$
and $H^2(X_+,\alpha)$ is the closure of $H^\infty(X_+,\alpha)$
with respect to the norm \eqref{4.6}.
\end{definition}
Note that the absolute value of a function from $H^2(X_+,\alpha)$
is single valued and $\alpha$ fixes, actually,  the ramification
of the argument of the function.

\smallskip

\noindent
{\bf Example}. As it was mentioned, the function $b$ is not single valued
but $|b(z)|$ is a single valued function. We define the character $\mu\in
\Gamma^*$ by
$$
b(\gamma z)=\mu(\gamma) b(z).
$$
Let $\gamma_j$ be  the contour, that starts at infinity (or any
other real point bigger than $1$), go in the upper half--plane to
the gap $(a_j,b_{j})$ and then go back in the lower half--plane to
the initial point. Assuming that
$b_0<...<a_j<b_j<a_{j+1}<...<a_0$, we have
$\mu(\gamma_j)=e^{-{2\pi i}\frac{g+1-j}{g+1}}$, equivalently
$\omega([b_j,a_{0}])=\frac {g+1-j} {g+1}$.

\smallskip

\noindent
{\bf Remark}. Note that the system of the above contours $\gamma_j$
is a generator of the free group $\Gamma^*(E)$. In other words
a character $\alpha$ is uniquely defined by the vector
$$
\begin{bmatrix}
\alpha(\gamma_1),&\alpha(\gamma_2),&\dots,&\alpha(\gamma_g)
\end{bmatrix}\in \bbT^g.
$$
This sets an one--to--one correspondence between $\Gamma^*(E)$ and $\bbT^g$.

\begin{proposition} Using the above definitions we get the orthogonal
decomposition
\begin{equation}\label{4.7}
  H^2(X_+)=\{1\}\oplus  H_0^2(X_+)
  =\{1\}\oplus b H^2(X_+,\mu^{-1}).
\end{equation}
\end{proposition}

Now we can iterate \eqref{4.7}. Let $k^\alpha$ be the reproducing kernel
of $H^2(X_+,\alpha)$ with respect to infinity, that is, the vector from
$H^2(X_+,\alpha)$,
which is uniquely defined by the condition
$$
\langle f, k^\alpha\rangle= f(\infty), \ \forall f\in H^2(X_+,\alpha).
$$
Then
\begin{equation*}
  H^2(X_+)=\{1\}\oplus \{b k^{\mu^{-1}}\}\oplus b^2H^2(X_+,\mu^{-2}),
\end{equation*}
and so on...

\begin{theorem}\label{4.5th}
Let $\alpha\in\Gamma^*$. The system
\begin{equation}\label{basis}
\{e_n^\alpha\}_{n\in \bbZ_+},
\quad e_n^\alpha:=
b^n K^{\alpha\mu^{-n}},
\ \  K^{\alpha\mu^{-n}}
:=\frac{k^{\alpha\mu^{-n}}}
{\Vert k^{\alpha\mu^{-n}}\Vert}
\end{equation}
forms an orthonormal basis of $H^2(X_+,\alpha)$. The same system
with $n\in \bbZ$ is an orthonormal basis in
$L^2(\partial X_+)$. Moreover, the multiplication operator by $z$
is a periodic Jacobi matrix with spectrum $E$.
\end{theorem}

Theorem \ref{4.5th} indicates a special importance of the functions
$k^{\alpha}$. They are very well studied \cite{Fay}. First of all, they
have analytic continuation (as multivalued functions) on the whole $X$,
so we can write
$k^{\alpha}(Z)$.

\begin{theorem} For every $\alpha\in \Gamma^*$
  the reproducing kernel $k^{\alpha}(Z)$ has on $X$ exactly
$g$ simple poles that do not depend on
$\alpha$ and
$g$ simple zeros. The divisor $D_+= \sum_{j=1}^g Z_j$
of zeros
\begin{equation}\label{4.9}
k^{\alpha}(Z_j)=0
\end{equation}
with the divisor of poles form the divisor
\begin{equation}\label{4.14s}
\div (k^{\alpha})= D_+- D_C
\end{equation}
that belongs to $D(E)$, moreover \eqref{4.14s} sets an  one--to--one
correspondence
between $D(E)$ and $\Gamma^*(E)$.
\end{theorem}
The functions $k^{\alpha}$ possess different representations, in
particular, in terms of theta--functions \cite{Mam}, and the map
$D\mapsto \alpha$ can be written explicitly in terms
of abelian integrals (the Abel map).

\bigskip

\noindent
{\bf Summary}. The three objects $J(E)$, $D(E)$ and $\Gamma^*(E)$
are equivalent. Both maps  $\Gamma^*(E)\to D(E)$
and $\Gamma^*(E)\to J(E)$
can be defined in terms of the reproducing kernels of the spaces
$H^2(X_+,\alpha)$,
$\alpha\in \Gamma^*(E)$. The first one is given by
\eqref{4.14s}. It associates to the given $k^{\alpha}(Z)$ the sets
of its zeros and poles (the poles are fixed and the zeros vary with $\alpha$).
The
matrix $J(\alpha)\in J(E)$ is defined as the matrix of the multiplication
operator by $z(Z)$ with respect to the basis \eqref{basis}:
\begin{equation}\label{jm}
z(Z)e^{\alpha}_s(Z)=p^\alpha_s e^{\alpha}_{s-1}(Z)+
q^\alpha_s e^{\alpha}_{s}(Z)+
p^\alpha_{s+1} e^{\alpha}_{s+1}(Z),\ Z\in X, \ s\in\bbZ.
\end{equation}

\bigskip

It's really easy to see that $J(\alpha)$
is periodic: just recall that $b^{g+1}$ is single valued, that is,
$\mu^{g+1}=1$, and therefore the spaces $H^2(X_+,\alpha)$
and $H^2(X_+,\alpha\mu^{-(g+1)})$ (and their reproducing kernels)
coincide. \bigskip

Now we can go back to the Renormalization Equation. Note that
$\pi$ acts naturally on $\Gamma(E)$:
$$
\pi\gamma =\{\pi(Z),\ Z\in\gamma\}\in \Gamma(\tilde E),
\quad \text{for}\ \gamma\in \Gamma(E).
$$
The map $\pi^*: \Gamma^*(\tilde E)\to \Gamma^*(E)$ is defined by
duality:
\begin{equation}\label{upch}
(\pi^* \tilde\alpha)(\gamma)= \tilde \alpha (\pi\gamma).
\end{equation}

\begin{theorem} \label{th4.7}
Let $T$, $T^{-1}:[-1,1]\to [-1,1]$,
be an expanding polynomial.
Let $\tilde J$ be a periodic Jacobi matrix with spectrum $\tilde E \subset
[-1,1]$, and therefore there exists a polynomial $\tilde U$ such that $\tilde
E=\tilde U^{-1} [-1,1]$ and a character $\tilde\alpha\in \Gamma^*(\tilde E)$
such that $\tilde J= J(\tilde \alpha)$. Then
\begin{equation}\label{solu1}
J:=J(\pi^*\alpha)
\end{equation}
is the periodic Jacobi matrix with spectrum $E= U^{-1} [-1,1]$, $U:=\tilde
U\circ T$, that satisfies the Renormalization Equation \eqref{t01}.
\end{theorem}

\begin{proof}
First we note, that for the operator multiplication
by $z(Z)$ in $L^2(\partial X_+)$, the operator multiplication
by $\tilde z(\tilde Z) $ in $L^2(\partial \tilde X_+)$,
the spectral parameter $z_0$ and the isometry
\begin{equation*}
(v f)(Z)=f(\pi(Z)),\quad v:L^2(\partial \tilde X_+)\to
L^2(\partial X_+),
\end{equation*}
we have
\begin{equation}\label{4.20s2}
\int_{\partial X_+} \frac 1{z_0-z(Z)} |(v f)(Z)|^2\, d\omega=
\int_{\partial \tilde X_+}\left(\frac  1 d
\sum_{\pi(Z)=\tilde Z}
\frac 1{z_0-z(Z)}\right) | f(\tilde Z)|^2\, d\tilde\omega.
\end{equation}
It is evident, that
\begin{equation*}
\frac  1 d
\sum_{T(y)=x}
\frac 1{z_0-y}=\frac{T'(z_0)/d}{T(z_0)-x}.
\end{equation*}
Thus
\begin{equation}\label{4.21s2}
v^*(z_0-z(Z))^{-1}v=(T'(z_0)/d)(T(z_0)-\tilde z(\tilde Z))^{-1}.
\end{equation}

It remains to show that $v$ transforms the basis vector
$$
\tilde e_n^{\tilde\alpha}=
\tilde b^n K^{\tilde\alpha\tilde\mu^{-n}}
$$
into
$$
  e_{nd}^{\pi^*\alpha}=
  b^{nd} K^{(\pi^*\tilde\alpha)\mu^{-n d}}=
(\tilde b^n\circ\pi) K^{\pi^*(\tilde\alpha\tilde\mu^{-n})}.
$$
Or, what is the same, that
$
  K^{\tilde\alpha}\circ\pi=K^{\pi^*\alpha}
$ for all $\tilde \alpha\in\Gamma^*(\tilde E)$.
Note that both functions are of norm one in the
same space $H^2(X_+,\pi^*\tilde \alpha)$, in particular, they have the
same character of automorphity $\pi^*\tilde \alpha\in \Gamma^*(E)$. Note,
finally, that the divisor
$$
\div (k^{\tilde\alpha}\circ\pi)= \pi^{-1} (\tilde D_+)-
\pi^{-1} (\tilde D_C),
$$
where
$\div (k^{\tilde\alpha})= \tilde D_+-\tilde D_C$,
belongs to $D(E)$, therefore $k^{\tilde\alpha}\circ\pi$ is the reproducing
kernel and the theorem is proved.
\end{proof}

To find all other solutions of \eqref{t01} let us look  a
bit more carefully at the above proof.

Note that the same identity \eqref{4.20s2} holds for any
isometry $v$ of the form
$$
vf=v_\theta f=\theta (f\circ\pi),
$$
where $\theta$ is a unimodular ($|\theta|=1$) function
on $\partial X_+$.

Concerning the second part of the proof, let us mention that
the set of critical points of $U$ splits in two sets:
$$
\{c_U\}=T^{-1}\{c_{\tilde U}\}\cup \{c_T\}.
$$
Correspondingly,
$$
\sum (C_U)_j=\sum_{k}
\sum_{\pi (C_{ U})_{k,j}=(C_{\tilde U})_k}(C_U)_{k,j}+\sum (C_T)_j,
$$
and the divisor of $k^{\tilde\alpha}\circ\pi$
consists of two parts, that one that depends on $\tilde \alpha$
$$
\pi^{-1}(\tilde D),
$$
and that part that corresponds to the critical points
of the polynomial $T$
$$
\{(C_T)_j\}_{j=1}^{d-1},
$$
since
$$
D=\div(k^{\tilde\alpha}\circ\pi)=
\pi^{-1}(\tilde D)+\sum_{j=1}^{d-1}(C_T)_j-
\pi^{-1}(\tilde D_C)-\sum_{j=1}^{d-1}(C_T)_j.
$$
Thus we can fix an arbitrary system of points $\{Z_{c,j}\}_{j=1}^{d-1}$ such
that
$z(Z_{c,j})$ belongs to the same gap in the spectrum $E$ as the critical point
$(c_T)_j$. If $\theta$ is the canonical product on $X$ with the divisor
$$
\div(\theta)=\sum_{j=1}^{d-1} Z_{c,j}-\sum_{j=1}^{d-1} (C_T)_{j},
$$
then $\theta k^{\tilde\alpha}\circ\pi$ is the reproducing kernel
simultaneously for all $\tilde \alpha\in\Gamma^*(\tilde E)$.
But to make $\theta$ unimodular
(zeros and poles are symmetric)
our choice is restricted just to
$Z_{c,j}=(C_T)_{j}$ or $Z_{c,j}=\tau (C_T)_{j}$. Note that
$\tau (C_T)_{j}- (C_T)_{j}$ is the devisor of the Complex Green function
$b_{(c_T)_j}$. In this way we arrive at

\begin{theorem}\label{th5.8} For an expanding polynomial $T$, and a
periodic Jacobi matrix
$\tilde J= J(\tilde \alpha)$, $\tilde\alpha\in\Gamma^*(\tilde E)$ as in Theorem
\ref{th4.7} there exist $2^{d-1}$ solutions of the Renormalization Equation
\eqref{t01}. Denote by $\mu_{(c_T)_j}$ the character generated by the Green's
function $b_{(c_T)_j}$, $ b_{(c_T)_j}\circ\gamma= \mu_{(c_T)_j}(\gamma)
b_{(c_T)_j}. $ Then these solutions are of the form
\begin{equation}\label{solu2}
J:=J\left(
\eta_\delta
\pi^*\tilde\alpha\right),
\quad  \eta_\delta:=\prod^{d-1}_{j=1}\mu_{(c_T)_j}^{\frac 1 2
(1+\delta_{(c_T)_j})},
\end{equation}
as before
\begin{equation*}
     \delta=\{\delta_{(c_T)_j}\},\quad \delta_{(c_T)_j}=\pm 1.
\end{equation*}
\end{theorem}

\begin{proof}
We define the isometry
$$
(vf)(Z)=\left(\prod^{d-1}_{j=1}
b_{(c_T)_j}^{\frac 1 2(1+\delta_{(c_T)_j})}(Z)\right)
f(\pi(Z))
$$
and then repeat the arguments of the proof of Theorem \ref{th4.7}.
\end{proof}

Concluding this section note that
the central part in the proof of Theorem \ref{lp}
(the claim
that the limit matrix has a form of the orthogonal sum)
also
can be reduced to an another well known fact from the theory of
Hardy spaces on Riemann surfaces. Namely to the statement
that $H^2$ is trivial, i.e.,
\begin{equation}\label{htrivial}
H^2(\bar \bbC\setminus E)=\{const\}
\end{equation}
in a domain of the form $\bar \bbC\setminus E$, where the Lebesgue measure of
$|E|=0$.

\begin{proof}[An alternative proof of \eqref{4.1}]
Since we can start with an arbitrary $\tilde J$, we  start with
a periodic matrix related to a certain $H^2(\tilde X_+)$,
e.g., with the matrix with constant coefficients
$\tilde J=\frac{S+S^*} 2$, $\tilde E=[-1,1]$. Then, under
inverse iterations of the polynomial $T$ according to Theorem
\ref{th4.7}, we will get  spaces of the same nature (i.e., the character
is trivial, equals one on every contour).
Let $J_n$ be the matrix with
spectrum $E_n=(T^n)^{-1}[-1,1]$. We have
\begin{equation*}
z(Z)e_1(Z,n)=p(n)_1 e_{0}(Z,n)+
q(n)_1 e_{1}(Z)+
p_{2} e_{2}(Z,n),\ Z\in X_n,
\end{equation*}
here $n$ is related to the number of iterations and the position
of the element of the matrix is fixed. Recall that
$e_{0}(Z,n)=1$ (the initial basic vector, see \eqref{4.7})
and we have, putting $Z=\infty$,
$$
p(n)_1=(zb_n)(\infty)K^{\mu_n^{-1}}(\infty,n).
$$
$(zb_n)(\infty)$ is the so called capacity of $E_n$, if it goes to zero
even better, in fact it does not, but in any case it is uniformly
bounded. Then,
assuming that $K^{\mu_n^{-1}}(\infty,n)$ does not go to zero, by compactness
arguments, we can find a subsequence
$$
(b_{n_j})(z)K^{\mu_{n_j}^{-1}}(z,n_j)
$$
that converges pointwise in the domain to a non--trivial holomorphic
function from $H^2(\bar \bbC\setminus E)$, $E=\lim E_n=\Julia(T)$, that equals
zero at $\infty$. But this contradicts to \eqref{htrivial}. Thus $p(n)_1\to 0$.
\end{proof}

Note that this proof is valid for expanding polynomials (we do not require that
$T$ is sufficiently hyperbolic). Note also the flip in notations of
the matrices'
elements: in this section a basis of holomorphic functions substitutes the
standard polynomial basis (instead of the multiplicity of the pole at infinity
we enlarge the multiplicity of zero). That is, $p_1$ in this section
is the same
as $p_0$ in Section 4 (we are just unable to enumerate the elements, related to
holomorphic functions, by negative integers).

\section{Concluding remarks}

Our concluding remarks concern basically other solutions of the
Renormalization Equation.

\subsection{The duality $\delta\mapsto-\delta$}
In Theorem \ref{mainth} we proved contractibility of
only one of the solutions of the renormalization equation corresponding to
$\delta=\delta_-$, but it means that at least one more solution has the same
property.
\begin{theorem}\label{th6.1}
The dual solution of the Renormalization Equation $J(\tilde J,-\delta)$,
  possesses the contractibility
property simultaneously with  $J(\tilde J,\delta)$.
\end{theorem}
It deals with the
following universal involution acting on  Jacobi matrices
\begin{equation}  \label{is6.1}
J\to J_\tau:=U_\tau J U_\tau, \quad\text{where}\ U_\tau
|l\rangle=|1-l\rangle.
\end{equation}
Obviously $VU_\tau=U_\tau S^{1-d}V$. Thus, having $J$ as a solution of
the renormalization equation corresponding to $\tilde J$ we have
simultaneously that $S^{d-1}J_\tau S^{1-d}$  solves the equation with
the initial $\tilde J_\tau$.  The following lemma describes which branch
corresponds
to which in this case.

\begin{lemma}
Let $J=J(\tilde J,\delta)$ then
\begin{equation}\label{is6.2}
   S^{d-1}J_\tau S^{1-d}=J(\tilde J_\tau,-\delta).
\end{equation}
\end{lemma}
\begin{proof}
We give a proof using the language of Sect. 5, so formally we prove the claim
only for periodic matrices.

Note that the involution
\eqref{is6.1} is strongly related to the standard involution $\tau$
\eqref{invol} on $X$.
Indeed, the function $K(\tau Z, \alpha)$ has the divisor
$$
\tau D_+ -\tau D_C=(\tau D_+ - D_C)-(\tau D_C-D_C),
$$
that is,
\begin{equation*}
K(\tau Z, \alpha)=\frac{K( Z, \beta)}{b_{c_1}(Z)\dots b_{c_{g}}(Z)},
\end{equation*}
and $\beta=\nu\alpha^{-1}$, where $\nu=\mu_{c_1} \dots   \mu_{c_g} $.
Due to this remark and the property $z(\tau Z)=z(Z)$ we have
\begin{equation}\label{is6.3}
(J(\alpha))_\tau=J(\nu\mu\alpha^{-1}).
\end{equation}

Now we apply \eqref{is6.3} to prove \eqref{is6.2}.  Let $\tilde
J_\tau=J(\tilde\alpha)$ with $\tilde\alpha\in \Gamma^*(\tilde X_+)$. Or,
in other words, $\tilde J=J(\tilde\mu\tilde\nu\tilde\alpha^{-1})$.
Then by \eqref{solu2}
\begin{equation*}
     J(\tilde
     J,\delta)=J(\eta_\delta\pi^*(\tilde\mu\tilde\nu\tilde\alpha^{-1})),
     \quad  \eta_\delta:=\prod^{d-1}_{j=1}
\mu_{(c_T)_j}^{\frac 1 2 (1+\delta_{(c_T)_j})}.
\end{equation*}
But $\pi^*\tilde\mu=\mu^d$ and $\pi^*(\tilde\nu)=\nu\eta_{\delta_+}^{-1}$
(just to look at the characters of the corresponding Blaschke products).
Thus, having in mind that $\eta_{\delta}  \eta_{-\delta}= \eta_{\delta_+}$,
we obtain
\begin{equation*}
    J(\tilde
    J,\delta)=J(\mu^d\nu\eta^{-1}_{-\delta}\pi^*(\tilde\alpha^{-1})).
\end{equation*}
Using again \eqref{is6.3} we get
\begin{equation*}
    (J(\tilde
    J,\delta))_\tau=J(\mu^{1-d}\eta_{-\delta}\pi^*(\tilde\alpha))
    =S^{1-d}J(\eta_{-\delta}\pi^*(\tilde\alpha))S^{d-1},
\end{equation*}
and the lemma and Theorem \ref{th6.1} are proved.
\end{proof}

Having two different contractive branches of solutions of the
renormalization equation, following \cite{Kn}, to an arbitrary sequence

\begin{equation*}
\epsilon=\{\epsilon_0,\epsilon_1,\epsilon_2\dots\},
\quad \epsilon_j=\delta_{\pm}.
\end{equation*}
we can associate a limit periodic
matrix $J$ with the spectrum on $\Julia(T)$. For a fixed sufficiently
hyperbolic
polynomial $T$, we define $J$ as the limit of
\begin{equation}\label{is6.4}
J_n:=J(
\eta_{\epsilon_0} \pi^*  \eta_{\epsilon_1} \dots   \pi^*\eta_{\epsilon_{n-1}}).
\end{equation}

\subsection{Other solutions of the Renormalization Equation and the
Ruelle operators}  We conjecture that actually all branches of
solutions of the renormalization equation are contractions for
sufficiently hyperbolic $T$. At least the previous remark looks as
a  quite strong indication in this direction: considering,
instead of initial $T$,  $T^2=T\circ T$ or its bigger powers, we
get, as in \eqref{is6.4}, several $\delta$'s,
  $ \eta_{\delta} =\eta_{\epsilon_0} \pi^*  \eta_{\epsilon_1} \dots
  \pi^*\eta_{\epsilon_{n-1}}$,
  possessing the contractibility property with respect to the polynomial $T^n$
and different
from $\delta_{\pm}$ (related to $(\pi^*)^n$).

  Similarly to \eqref{is4.4},  \eqref{is4.5}
we formulate
\begin{conjecture}
     Let $T(z)$ be an expanding polynomial and let
     $T'(z)=A_1(z) A_2(z)$ be an arbitrary (polynomial) factorization
     of the derivative. Denote by $\sigma_{1,2}$,  the
     (nonnegative)
     eigen--measures, corresponding to the Ruelle operators
     \begin{equation}\label{}
     (L_{A_i}f)(x)=\sum_{T(y)=x}\frac {f(y)}{A_i(y)^2},
\end{equation}
i.e.,  $L^*_{A_i}\sigma_i=\rho_i  \sigma_i $. Finally let
$J_{1,2}$ be the one-sided Jacobi matrices associated with
$\sigma_{1,2}$. Then the
block matrix $J=J_-\oplus J_+$ with $J_-=J_1$ and $J_+=J_2$ is limit periodic.
\end{conjecture}
Note that by the same reason as above the conjecture holds true,
   say, for $T^2(z)$ and $A_1(z)=T'(z)$, $A_2= T'(T(z))$.

\subsection{Shift transformations with the Lipschitz property}
We say that the direction $\eta\in \Gamma^*$ has   the Lipschitz property
with a constant $C(\eta)$ if for all   $\alpha, \beta\in \Gamma^*$
\begin{equation}\label{}
     \Vert J(\eta\alpha)-J(\eta\beta)\Vert\le   C(\eta)
     \Vert J(\alpha)-J(\beta)\Vert.
\end{equation}
Then, one can get the contractibility of the map $\eta\pi^*$ in two steps:
\begin{equation}
\begin{split}
     \Vert J(\eta\pi^*\tilde\alpha)-J(\eta\pi^*\tilde\beta)\Vert &\le   C(\eta)
     \Vert J(\pi^*\tilde\alpha)-J(\pi^*\tilde\beta)\Vert   \\
     &\le C(\eta)\varkappa
     \Vert J(\tilde\alpha)-J(\tilde\beta)\Vert.
     \end{split}
  \end{equation}

Note, that in fact the situation is a bit more involved because we
should be able to
compare Jacobi matrices with different spectral sets, for example, when
$E_i=T^{-1}\tilde E_i$, $\tilde E_1\not=\tilde E_2$. But we just wanted to
indicate the general idea, in particular, for directions $\eta_{\delta}$ of the
form \eqref{solu2} such a comparison is possible. Of course, for our goal the
constant $C(\eta)$ should be uniformly bounded when we increase the level of
sufficient hyperbolicity of $T$ making $\varkappa$ smaller.

However the key point of this remark (this way of proof) is that,
actually, {\it we do not need to constrain ourselves   by the
form of the vector $\eta$}. Combining a  ``Lipschitz" shift by $\eta$ (the
direction is restricted just by this property) with a sufficiently
contractive pull--back $\pi^*$ we arrive at an iterative process that produces
a limit periodic Jacobi matrix with the spectrum on the same
$\Julia(T)$. In the next
subsection we give examples of directions with the required property, see
Corollary \ref{cordbr}.

We do not
have a proof of the Lipschitz property of
$\eta_\delta$'s, but there is a good chance to generalize the result
of the next
subsection in a way that at least some of the directions
$\eta_\delta$ will be also
available.

Finally, we would be very interested to know, whether there is in
general a relation
between the form of the ``weight" vector $\eta$ and the corresponding
weights of the
Ruelle operators (if any exists).

\subsection{Quadratic polynomials and
the Lipschitz property of the Darboux transform}
Consider the simplest special case $T(z)=\rho(z^2-1)+1$, $\rho>2$.
Note that  the spectral set
$E=T^{-1}\tilde E$
is symmetric, moreover
the matrix related to $H^2(\pi^*\tilde\alpha)$ has zero main diagonal
(as well as a one--sided matrix related to a symmetric measure).
Now we introduce a decomposition of $H^2(\pi^*\tilde\alpha)$
which is
very similar to the standard decomposition into even and odd functions.

We define the two--dimensional vector--function representation of $f\in
H^2(\pi^*\tilde\alpha)$
\begin{equation}\label{30s36}
f\mapsto\frac 1{\sqrt{2}}
\begin{bmatrix} f(Z_1(\tilde Z)) \\ f(Z_2(\tilde Z))
\end{bmatrix}\mapsto
\begin{bmatrix} g_1(\tilde Z)\\ g_2(\tilde Z)
\end{bmatrix},
\end{equation}
where
\begin{equation*}
\begin{bmatrix} g_1(\tilde Z)\\ g_2(\tilde Z)
\end{bmatrix}=
\frac 1 2
\begin{bmatrix} f(Z_1(\tilde Z)) +f(Z_2(\tilde Z))\\
f(Z_1(\tilde Z))-f(Z_2(\tilde Z))
\end{bmatrix},
\end{equation*}
the first component, in a sense, is even and the second is odd.
To be more precise, let us describe analytical properties of this
object in $\partial \tilde X_+$.

Note that
due to
$$
\int_{\partial X_+}|f|^2 d\omega=
\int_{\partial \tilde X_+}\frac 1 2\sum_{\pi(Z)=\tilde Z}|f|^2 d
\tilde\omega
$$
metrically it is of $L^2$ with respect to $\tilde\omega$,
moreover the transformation is norm--preserved.

It is evident that the function
$g_1$
belongs to $H^2(\tilde X_+, \tilde\alpha)$. Consider the second function.
Note that the critical points of $T$ are zero and infinity.
For a small circle $\gamma$ around the point
$T(0)=-\rho+1$ we have
$g_2\circ\gamma=-g_2$
and the same property for a contour $\gamma$ that surrounds infinity. Let us
introduce
$$
\Delta^2:= \tilde b_{T(0)} \tilde b.
$$
Note that for the above contours we have $\Delta\circ\gamma=- \Delta$.
We are going to represent $g_2$ in the form $g_2=\Delta\hat g_2$ and to claim
that
$\hat g_2$ has nice automorphic properties in $\tilde X_+$. Let us note that
\begin{equation*}
\tilde b\frac{\tilde z-T(0)}{\tilde b_{T(0)}}=
\tilde b^2\frac{\tilde z-T(0)}{\Delta^2}
\end{equation*}
is an outer function in the domain $\bar{\bbC}\setminus \tilde E\simeq
\tilde X_+$.
So, the square root of this function is well defined. We put
\begin{equation}\label{sqroot}
\tilde b \phi:=\sqrt{
\tilde b^2\frac{\tilde z-T(0)}{\rho \Delta^2}}
\end{equation}
and denote by $\tilde \eta$ the character generated by $ \phi$,
$ \phi\circ\gamma= \eta(\gamma) \phi$. Thus
\eqref{sqroot} reduces the ramification of the function
$\Delta$ to the function $\phi$, which is well defined in the domain,
and to the
elementary function $\sqrt{\tilde z-T(0)}$.

\begin{theorem}
The transformation $f\mapsto g_1\oplus \hat g_2$ given by
\eqref{30s36} is a unitary map from $H^2(\pi^*\tilde\alpha)$
to $H^2(\tilde\alpha)\oplus H^2(\tilde\alpha\tilde\eta)$.
Moreover with respect to this representation
\begin{equation}\label{12o40}
z f\mapsto
\begin{bmatrix}0& \bar\phi\\
\phi &0
\end{bmatrix}\begin{bmatrix}
g_1\\ \hat g_2
\end{bmatrix}
\end{equation}
and
$$
v_+f\mapsto f\oplus 0,\quad f\in H^2(\tilde\alpha),
$$
where the isometry $v_+:H^2(\tilde\alpha)\to H^2(\pi^*\tilde\alpha)$
is defined by \eqref{encl2}.
\end{theorem}

\begin{proof}
By the definition of $\Delta$ we have
\begin{equation}\label{12o38}
g_2=\Delta \hat g_2,\
\text{where}\ \hat g_2\in H^2(\tilde\alpha\tilde\eta).
\end{equation}
Further,
since
$$
  z_{1,2}=\pm\sqrt{\frac{\tilde z-T(0)}{\rho}},
$$
we have, say for the second component,
\begin{equation}\label{12sept38}
\frac{1}{\Delta}\frac{(zf)(Z(\tilde Z_1))- (zf)(Z(\tilde Z_2))}{2}
=\sqrt{\frac{\tilde z-T(0)}{\rho\Delta^2}}
\frac{f(Z(\tilde Z_1))+ f(Z(\tilde Z_2))}{2}=
\phi g_1.
\end{equation}
Since on the boundary of the domain
$$
\phi^2\Delta^2=\frac {\tilde z-T(0)}{\rho}=|\phi|^2
$$
(the second expression is positive on $\partial \tilde X_+$)
we have
\begin{equation}\label{12o41}
\phi\Delta^2=\overline{\phi} \quad {\rm on}\ \tilde E.
\end{equation}
Using this relation, similarly to \eqref{12sept38}, we prove the identity
of the first components in \eqref{12o40}.

\end{proof}

\begin{theorem}
The multiplication operator
$\phi: L^2(\partial \tilde X_+)\to L^2(\partial \tilde X_+)$
with respect to the basis systems \eqref{basis} related to $\tilde\alpha$ and
$\tilde \eta\tilde \alpha$, respectively, is a two diagonal matrix $\Phi$.
Moreover,
\begin{equation}\label{12o42}
\Phi^*\Phi= \frac{J(\tilde \alpha)-T(0)}{\rho},
\quad
\Phi\Phi^*= \frac{J(\tilde\eta\tilde \alpha)-T(0)}{\rho}.
\end{equation}
In other words, the transformation
$J(\tilde \alpha)\mapsto J(\tilde\eta\tilde \alpha)$ is the Darboux
transform.
\end{theorem}

\begin{proof} First of all $\phi$ is a character--automorphic function with the
character $\tilde\eta$ with a unique pole at infinity ($\tilde b\phi$ is an
outer function). Therefore the multiplication operator acts from $\tilde
bH^2(\tilde\alpha \tilde\mu^{-1})$ to $H^2(\tilde\eta\tilde\alpha)$. Therefore,
the operator $\Phi$ has only one non--trivial diagonal above the main diagonal.
The adjoint operator has the symbol $\overline \phi$. According to
\eqref{12o41}
it has holomorphic continuation from the boundary inside the domain. Thus
$\Phi^*$ is a lower triangular matrix. Combining these two facts we get that
$\Phi$ has only two non--trivial diagonals. Then, just comparing symbols of
operators on the left and right parts of \eqref{12o42}, we prove these
identities.
\end{proof}

\begin{corollary}\label{cordbr}
    Let $\tilde J_{1,2}$ be periodic Jacobi matrices with the
spectrum on $[-1,1]$. Let
    $\Drb(\tilde J_{1,2},\rho)$ be their Darboux transforms. Then
\begin{equation}\label{drx}
\Vert\Drb(\tilde J_{1},\rho)-\Drb(\tilde J_{2},\rho)\Vert
\le C(\rho)
\Vert\tilde J_{1}-\tilde J_{2}\Vert.
\end{equation}
\end{corollary}
\begin{proof}
For the given $\tilde J_{1,2}$ we define $J_{1,2}$ via the quadratic polynomial
$T(z)=\rho(z^2-1)+1$. Being decomposed into even and odd indexed subspaces they
are of the form
\begin{equation}\label{drx2}
J_{1,2}=\begin{bmatrix}
0&\Phi^*_{1,2}\\
\Phi_{1,2}& 0
\end{bmatrix}.
\end{equation}
Due to the main theorem, that gives the uniform estimate for
$\Vert J_1-J_2\Vert$, we have
\begin{equation}\label{drx}
\Vert\Phi_{1}-\Phi_{2}\Vert
\le \kappa(\rho)
\Vert\tilde J_{1}-\tilde J_{2}\Vert,
\end{equation}
with $\kappa(\rho)=\frac{C}{\rho-2}$, $C$ is an absolute constant.
Using \eqref{12o42} we get \eqref{drx} with
$C(\rho)=\frac{2\rho C}{\rho-2}$.
\end{proof}

\section{Appendix}

Here we recall some basic facts on two--sided Jacobi matrices.
Let $J$ define a bounded selfadjoint operator on
$l^2(\bbZ)$. The resolvent matrix--function
is defined by the relation
\begin{equation}
W(z)=W(z,J)=\begin{bmatrix}
\langle 0|(J-z)^{-1}|0\rangle& \langle 0|(J-z)^{-1}|1\rangle\\
\langle 1|(J-z)^{-1}|0\rangle& \langle 1|(J-z)^{-1}|1\rangle
\end{bmatrix}.
\end{equation}

This matrix--function has an integral representation
\begin{equation}
W(z)=\int\frac{d\sigma(x)}{x-z}
\end{equation}
with $2\times 2$ matrix--measure having a compact support on $\bbR$.
$J$ is unitary equivalent to the multiplication operator
by an independent variable on
\begin{equation}
L^2_\sigma=\left\{
f=\begin{bmatrix} f_0(x)\\ f_1(x)\end{bmatrix}:
\int f^* d\sigma f<\infty
\right\},
\end{equation}
moreover, under this unitary mapping from $l^2\to L^2_\sigma$ we have
\begin{equation}
|0\rangle\mapsto\begin{bmatrix} 1\\ 0\end{bmatrix},
\quad
|1\rangle\mapsto\begin{bmatrix} 0\\ 1\end{bmatrix}.
\end{equation}

Let $r_-(z)=r_-(z,0)$, $r_+(z)=r_+(z,1)$ be resolvent functions
of $J_-=J_-(0)$ and $J_+=J_+(1)$ respectively (see \eqref{4aug22}),
and $\sigma_{\pm}$ be the corresponding (scalar!)
spectral measures.
Then
\begin{equation}\label{ap4.5}
W(z)=\begin{bmatrix} r_-^{-1}(z)& p_1\\
p_1& r_+^{-1}(z)
\end{bmatrix}^{-1}.
\end{equation}

Recall that according to our notation $p_1 Q_d$ is the orthonormal
polynomial of the degree $d-1$
for the measure $\sigma_+$ and $p_1 R_d$ is the related polynomial of
the second kind:
\begin{equation}
p_1 R_d(z)=\int d\sigma_+(x)\frac{p_1 Q_d(x)-p_1 Q_d(z)}{x-z}.
\end{equation}
In this notations
\begin{equation}
|d\rangle\mapsto\cE_d(x):=
\begin{bmatrix}
-p_1^2 R_d(x)
\\ p_1 Q_d(x)
\end{bmatrix},
\end{equation}
moreover
\begin{equation}\label{ap4.8}
\begin{bmatrix}
0
\\ p_1 R_d(x)
\end{bmatrix}=
\int d\sigma(x)\frac{\cE_d(x)-\cE_d(z)}{x-z}.
\end{equation}

\begin{lemma}
Let
\begin{equation}
\cF(x):=\begin{bmatrix} 1&
-p_1^2 R_d(x)
\\ 0 &  p_1 Q_d(x)
\end{bmatrix},
\quad
\cG(x):=\begin{bmatrix} 0&
0
\\ 0 &  p_1 R_d(x)
\end{bmatrix},
\end{equation}
Then
\begin{equation}\label{ap4.10}
\begin{bmatrix}
\langle 0|(J-z)^{-1}|0\rangle& \langle 0|(J-z)^{-1}|d\rangle\\
\langle d|(J-z)^{-1}|0\rangle& \langle d|(J-z)^{-1}|d\rangle
\end{bmatrix}=\cF^*(\bar z) W(z) \cF(z)+
\cF^*(\bar z)\cG(z).
\end{equation}
\end{lemma}
\begin{proof}
This is a standard trick from the theory of orthogonal polynomials.
Due to the unitary mapping onto $L^2_\sigma$, equivalently we have to
calculate
\begin{equation}
\int \frac{\cF(x)^*d\sigma(x) \cF(x)}{x-z}.
\end{equation}
Therefore, using orthogonality
and \eqref{ap4.8}, we continue
\begin{equation}
\begin{split}
=&\int \frac{\cF(x)^* - \cF(\bar z)^*}{x-z}d\sigma(x) \cF(x)+
\cF(\bar z)^*\int \frac{d\sigma(x) \cF(x)}{x-z}\\
=&\cF(\bar z)^*\int d\sigma(x)\frac{ \cF(x)-\cF(z)}{x-z}
+\cF(\bar z)^*\int \frac{ d\sigma(x)}{x-z}\cF(z)\\
=&\cF^*(z)\cG(z)+\cF^*(\bar z) W(z) \cF(z).
\end{split}
\end{equation}
\end{proof}

\begin{corollary}\label{cor4.2} Combining \eqref{ap4.5} with
\eqref{ap4.10} we get \eqref{re0} from
\begin{equation*}
\begin{bmatrix}
\langle 0|(J-z)^{-1}|0\rangle& \langle 0|(J-z)^{-1}|d\rangle\\
\langle d|(J-z)^{-1}|0\rangle& \langle d|(J-z)^{-1}|d\rangle
\end{bmatrix}=\frac{
\begin{bmatrix}
\langle 0|(\tilde J-T(z))^{-1}|0\rangle& \langle 0|(\tilde
J-T(z))^{-1}|1\rangle\\
\langle 1|(\tilde J-T(z))^{-1}|0\rangle&
\langle 1|(\tilde J-T(z))^{-1}|1\rangle
\end{bmatrix}}{T'(z)/d},
\end{equation*}
which is a part of the Renormalization Equation.
\end{corollary}
\begin{proof}
A straightforward computation.
\end{proof}

\begin{flushleft}
address: \\[.1cm]
Alexander Volberg\\
Department of Mathematics \\
Michigan State University \\
East Lansing, Michigan 48824, USA \\
volberg@math.msu.edu\\
\end{flushleft}

\begin{flushleft}
Address: \\[.1cm]
Franz Peherstorfer\\
Peter Yuditskii\\
Abteilung f\"ur Dynamische Systeme \\
und Approximationstheorie\\
Institut f\"ur Analysis \\
J. Kepler Universit\"at Linz\\
Linz, Austria A-4040
\end{flushleft}


\markboth{}{}

\begin{thebibliography}{99}

\bibitem{AvS} J.~Avron, B.~Simon, {\em Singular continuous spectrum
for a class of almost periodic Jacobi matrices} Bull. AMS {\bf 6}
(1982), 81--85.


\bibitem{BGH} M.~F.~Barnsley, J.~S.~Geronimo, A.~N.~Harrington,
{\em Almost periodic Jacobi matrices associated with Julia sets for
polynomials}, Comm. Math. Phys. {\bf 99} (1985), no. 3, 303--317.

\bibitem{BBM} J.~Bellissard, D.~Bessis, P.~Moussa, {\em Chaotic states
of almost periodic Schr\"odinger operators}, Phys. Rev. Lett. {\bf 49}
(1982), no. 10, 701--704.

\bibitem{BS} J.~Bellissard, B.~Simon, {\em Cantor spectrum for almost
Mathieu equation} J. Funct. Anal., {\bf 48} (1982), no. 3, 408--419.

\bibitem{Bo} R.~Bowen, {\em  Equilibrium States and the Ergodic Theory of
Anosov Diffeomorphisms}
Lecture Notes in Mathematics, v. 470, Springer-Verlag,
Berlin-Heidelberg-New York, 1975.


\bibitem{Fay}
J. ~Fay,
\textit{Theta Functions on Riemann Surfaces}, Lecture Notes in
Mathematics, Vol. 352, Springer-Verlag, New York/Berlin, 1973.


\bibitem{H}  J.~Herndon,
{\em Limit periodicity of sequences defined by certain recurrence
relations; and Julia sets},
Ph.D. thesis, Georgia Institute of Technology, 1985.


\bibitem{Kn}
O. ~Knill,
{\em Isospectral deformations of random Jacobi operators}.
Comm. Math. Phys. 151 (1993), no. 2, 403--426.

\bibitem{Mam}Ê
D. ~Mamford,  {\it Lectures on
theta-functions}. (Russian)  Translated
from the English by D. Yu. Manin. Translation edited and with a preface
by Yu. I. Manin. With appendices by Khirosi Umemura [Hiroshi Umemura]
and Takakhiro Shiota [Takahiro Shiota]. ``Mir'', Moscow, 1988. 448 pp.

\bibitem{SYu} M. ~Sodin, P. ~Yuditski,
{\em The limit-periodic
finite-difference operator on $l_2(\bbZ)$ associated with
iterations of quadratic polynomials}. J. Statist. Phys. 60 (1990), no.
5-6, 863--873.

\bibitem{VY} F. ~Pehersorfer, A. ~Volberg, P. ~Yuditskii,
{\em
Two weight
Hilbert transform and Lipschitz
property of Jacobi matrices associated to hyperbolic polynomials,}
submitted.


\end{thebibliography}
\end{document}